\documentclass{article}
\newtheorem{Theorem}{Theorem}

\newtheorem{Corollary}[Theorem]{Corollary}

\begin{document}
\title{Reaching Majority Agreement in~a~Disconnected~Network}
\author{
Chris Dowden \\
\\
Graz University of Technology \\
\\
dowden@math.tugraz.at 
}
\maketitle
\setlength{\unitlength}{1cm}

\begin{abstract}
We investigate the problem of reaching majority agreement in a disconnected network.
We obtain conditions under which such an agreement is certainly possible/impossible,
and observe that these coincide in the ternary case.
\end{abstract}

\section{Introduction}

Suppose various processors in a network wish to reach agreement on a particular decision.
In traditional agreement problems
(as first studied in \cite{Pease}),
it is usually assumed that the network is connected
(but perhaps suffering from faults in unknown locations),
and the aim is to achieve unanimous agreement.
Additionally,
work has also been carried out on the topic of `almost-everywhere' agreement
(see, for example, \cite{Dwork}).
In this paper,
we instead allow networks that are disconnected,
and the aim is to design a protocol that will nevertheless enable the processors to reach majority agreement.

We assume that all processors know the overall network and the protocol.
Each processor $i$ is then given an input value $v_{i}$,
but is only allowed to communicate with the other processors in its component.
After as much communication within each component as desired,
a successful protocol must then terminate with the processors having reached a majority agreement satisfying the following two requirements: \\
(i) a strict majority of the processors must agree on a common output value $v$; \\
(ii) if the input to all processors is $v_{c}$,
then $v$ must be $v_{c}$.

The binary case 
(when there are only two possible input values)
is fairly trivial,
so we shall concentrate on the general case.
In Theorem~\ref{possible}/Corollary~\ref{corollary},
we shall provide a simple protocol that works successfully when the network contains a large component;
in Theorem~\ref{impossible},
we shall then derive conditions for which no such protocol is possible.

\section{Results}

As mentioned,
we start with a straightforward argument that establishes conditions under which majority agreement protocols are possible.
In the resulting corollary,
we shall then simplify the bound obtained.

\begin{Theorem} \label{possible}
Suppose there are $k$ potential input values,
and let $G$ be a network whose largest component has $h$ processors,
where $h$ satisfies $h + \lfloor \frac{|G|-h}{k} \rfloor > \frac{|G|}{2}$.
Then there exists a successful majority agreement protocol.
\end{Theorem}
\textbf{Proof}
Let $G$ satisfy the conditions of the theorem,
and (without loss of generality)
let $1,2, \ldots, k$  be the potential inputs.
One successful protocol is simply to have all processors in the largest component
(with some pre-agreed rule for ties)
output whatever is the most common input in that component
(again using some pre-agreed rule for ties),
while also choosing $\lfloor \frac{|G|-h}{k} \rfloor$ of the other processors to always output $1$
(regardless of inputs),
choosing $\lfloor \frac{|G|-h}{k} \rfloor$ to always output $2$,
and so on.
The inequality $h + \lfloor \frac{|G|-h}{k} \rfloor > \frac{|G|}{2}$
then guarantees that whichever value is the most common input in the largest component will achieve a strict majority of the outputs.
\setlength{\unitlength}{.25cm}
\begin{picture}(1,1)
\put(0,0){\line(1,0){1}}
\put(0,0){\line(0,1){1}}
\put(1,1){\line(-1,0){1}}
\put(1,1){\line(0,-1){1}}
\end{picture}

\begin{Corollary} \label{corollary}
Suppose there are $k$ potential input values,
and let $G$ be a network whose largest component has $h$ processors,
where $h$ satisfies $\frac{|G|-h}{k} \in \mathbf{Z}$ and $h > \left( \frac{k-2}{2(k-1)} \right) |G|$.
Then there exists a successful majority agreement protocol.
\end{Corollary}

We now proceed with our main result,
which establishes conditions under which no majority agreement protocol is possible.
Note that the bound given for the largest component matches the converse result from Corollary~\ref{corollary}.

\begin{Theorem} \label{impossible}
Suppose there are $k \geq 3$ potential input values,
and let $G$ be a network whose largest component has at most $\left( \frac{k-2}{2(k-1)} \right) |G|$ processors
and whose two largest components together have at most $\frac{|G|}{2}$ processors.
Then no majority agreement protocol is possible.
\end{Theorem}
\textbf{Proof}
Let $G$ satisfy the conditions of the theorem,
and let $H$ and $J$ denote the largest and second largest components, respectively.
Thus,
we have $|H| \leq \left( \frac{k-2}{2(k-1)} \right) |G|$
and $|H| + |J| \leq \frac{|G|}{2}$.

Without loss of generality,
let $1,2, \ldots, k$ be the potential inputs,
and let us suppose
(aiming for a contradiction)
that there does exist a successful majority agreement protocol.
We shall use this protocol throughout the remainder of the proof.

From among all sets of inputs that result in $1$ being the majority output
(there is at least one such set,
since we could take all the inputs to be $1$),
let us look at one which minimises the number of $1$'s in the output.

By definition,
the number of $1$'s in the output produced from this set must be greater than $\frac{|G|}{2}$.
Note that this then implies that there must exist some value $i \in \{2,3, \ldots, k\}$
such that less than $\frac{|G|}{2(k-1)}$ of the processors output $i$.
Without loss of generality,
let us take $i$ to be $3$.

We know that if we were to change all inputs to $3$,
then $3$ would be the majority output rather than $1$.
Note that to achieve this overall reduction in the number of $1$'s in the output,
there must exist some component $G_{1}$ for which changing all inputs to $3$ in just this particular component
will reduce the number of $1$'s in the output
(this observation uses the fact that the output of a processor can only depend
on the inputs of the processors in its component,
and we shall implicitly use this independence of different components throughout the proof).

By our minimality condition,
it can only be that $1$ is then no longer the majority output.
Note also that the number of $3$'s in the output is still at most $\frac{|G|}{2(k-1)} + |G_{1}| \leq \frac{|G|}{2(k-1)} + |H| \leq \frac{1}{2}|G|$,
and so $3$ is also not the majority output.
Thus, without loss of generality,
let $2$ be the new majority output,
and observe that the number of $2$'s in the output must previously have already been greater than $\frac{|G|}{2} - |G_{1}|$.
Consequently,
for all values $u \in \{3,4, \ldots, k\}$,
the number of $u$'s in the output must have been less than $|G_{1}|$.

Let us return to this original set of inputs.
For $u \notin \{1,2\}$,
we have just established that the number of $u$'s in the output is less than $|G_{1}|$,
and so the inequality $|H| + |J| \leq \frac{|G|}{2}$
then implies that it is also not possible to change all inputs to $3$ in any alternative component (instead of $G_{1}$)
in such a way that $u$ becomes the majority output.
Hence,
it follows from our minimality condition
that it is not possible to change all inputs to $3$ in any single component
in such a way that the number of $1$'s in the output and the number of $2$'s in the output both reduce.
We shall use this important observation later.

Let us use $l$ to denote the number of components of $G$.
Working component by component,
and thus taking $l$ steps in total,
we shall now change all inputs in all components to $3$.
We shall commence in Step $1$ with component $G_{1}$,
and our procedure will then involve choosing a careful order $G_{2}, G_{3}, \ldots, G_{l}$ for the remaining components.
We shall show that it is possible to choose this ordering so that 
the output after Step $j$ always satisfies the following three inequalities: 
\begin{eqnarray*}
\max \{ \textrm{number of $1$'s, number of $2$'s} \} & > & \frac{|G|}{2}; \\
\min \{ \textrm{number of $1$'s, number of $2$'s} \} & > & \frac{|G|}{2} - \max_{i \leq j}|G_{i}|; \\
\textrm{for all } u \notin \{1,2\}, \textrm{ number of $u$'s} & < & \max_{i \leq j}|G_{i}|. 
\end{eqnarray*}

The proof is by induction.
Note that the base case
(i.e.~after Step $1$)
follows from the fact that we know $2$ to be the majority output,
together with the observation that the number of $1$'s in the output cannot have decreased by more than $|G_{1}|$ during Step $1$
and so must still be greater than $\frac{|G|}{2} - |G_{1}|$
(thus leaving less than $|G_{1}|$ processors for the other output values).
Let us now suppose that the desired inequalities all hold at the end of Step $r$ for some $r \in \{1, \ldots, l-1 \}$,
and let us consider Step $r+1$.

Let $v$ denote the value (either $1$ or $2$)
that is in the majority at the end of Step $r$,
and let $w$ denote whichever of $1$ or $2$ is not in the majority.
In Step $r+1$,
let us find a component $G_{r+1}$ for which changing all the inputs in $G_{r+1}$ to $3$ decreases the number of $v$'s in the output
(there must exist such a component,
since changing all the inputs in the whole of $G$ to $3$ would result in $3$ becoming the majority output instead of $v$).

Consider $u \notin \{1,2\}$.
By the induction hypothesis,
the number of $u$'s in the output after Step $r$ was less than $\max_{i \leq r}|G_{i}|$,
and this number can only have increased by at most $|G_{r+1}| \leq \frac{|G|}{2} - \max_{i \leq r}|G_{i}|$ during Step $r+1$.
Thus,
the majority output can still only be either $1$ or $2$,
and so the output after Step $r+1$ still satisfies
$\max \{ \textrm{number of $1$'s, number of $2$'s} \} > \frac{|G|}{2}$.

Now recall our earlier important observation
that it is not possible to change the inputs in any single component
in such a way that the number of $1$'s in the output and the number of $2$'s in the output both reduce.
Since $G_{r+1}$ was chosen so that the number of $v$'s in the output reduces,
it then follows that the number of $w$'s in the output cannot have reduced
and so must still be greater than $\frac{|G|}{2} - \max_{i \leq r}|G_{i}|$.
Note also that the number of $v$'s in the output must still be greater than
$\frac{|G|}{2} - |G_{r+1}|$.
Hence, putting these bounds for $w$ and $v$ together,
we thus conclude that the output after Step $r+1$ must certainly satisfy
$\min \{ \textrm{number of $1$'s, number of $2$'s} \} > \frac{|G|}{2} - \max_{i \leq r+1}|G_{i}|$.

By subtraction,
we then also find that for all $u \in \{1,2\}$,
the number of $u$'s in the output after Step $r+1$ 
must certainly be less than $\max_{i \leq r+1}|G_{i}|$.

Thus, by induction,
our inequalities all hold as stated.
Observe that,
in particular,
this implies that the number of $3$'s in the output after Step $l$ is less than $|H|$.
Hence, we have achieved our desired contradiction,
since after Step $l$ all inputs will now be $3$,
and so $3$ should be the majority output.
\phantom{qwerty}
\setlength{\unitlength}{.25cm}
\begin{picture}(1,1)
\put(0,0){\line(1,0){1}}
\put(0,0){\line(0,1){1}}
\put(1,1){\line(-1,0){1}}
\put(1,1){\line(0,-1){1}}
\end{picture} \\

For the ternary case when $k=3$,
we note that the conditions in the statement of Theorem~\ref{impossible}
simplify to just a bound (of $\frac{|G|}{4})$ on the largest component.
By Corollary~\ref{corollary},
we know that this bound is tight,
and so the ternary case is thus solved.
For larger values of $k$,
it still remains to deal with the extra condition imposed in Theorem~\ref{impossible}
on the combined size of the two largest components.

\section{Concluding Remarks}

In this paper,
we have studied the topic of how a collection of processors in a disconnected network $G$ 
can attempt to reach majority agreement when choosing from a set of $k$ possible values.

For the ternary case when $k=3$,
we have deduced that it is necessary and sufficient
(subject to certain integrality requirements)
for more than a quarter of the processors to be in the same component.
For the general case,
we have observed a successful procedure for when the largest component contains more than
$\left( \frac{k-2}{2(k-1)} \right) |G|$ processors,
and shown that no such protocols exist beneath this bound
as long as the combined size of the two largest components is at most $\frac{|G|}{2}$.
The necessity of this extra condition is left unresolved.

\section*{Acknowledgments}
This work was written at Royal Holloway and Bedford New College,
University of London,
as part of the EU-funded `Internet of Energy for Electric Mobility' project.

\end{document}